\documentclass[prl,twocolumn,superscriptaddress,showpacs]{revtex4}

\usepackage{mathtools,wasysym,latexsym,bbold,comment}
\usepackage{graphicx,amssymb,amsmath,amsfonts,psfrag,color}
\usepackage[colorlinks=true, urlcolor=blue, linkcolor=red, citecolor=blue, pdftex]{hyperref}

\newcommand{\la}{\langle}
\newcommand{\ra}{\rangle}
\newcommand{\lnb}{\left(}
\newcommand{\rnb}{\right)}
\newcommand{\nn}{\nonumber}

\begin{document}

\title{A fractionalised ``$\mathbb{Z}_2$'' classical Heisenberg spin liquid}
\author{J. Rehn}
\affiliation{Max-Planck-Institut f\"ur Physik komplexer Systeme, 01187 Dresden, Germany}
\author{Arnab Sen}
\affiliation{Department of Theoretical Physics, Indian Association for the Cultivation of Science, Kolkata 700032, India}
\author{R. Moessner}
\affiliation{Max-Planck-Institut f\"ur Physik komplexer Systeme, 01187 Dresden, Germany}
\date{\today}

\begin{abstract}
Quantum spin systems are by now known to exhibit a large number of
different classes of spin liquid phases.  By contrast, for
{\em  classical} Heisenberg models, only one kind of fractionalised
spin liquid phase, the so-called Coulomb or $U(1)$ spin liquid, has
until recently been identified: this exhibits
algebraic spin correlations and impurity moments, `orphan spins',
whose size is a fraction of that of the underlying microscopic
degrees of freedom.  Here, we present two Heisenberg models exhibiting
fractionalisation in combination with exponentially decaying correlations.
These can be thought of as a classical continuous spin version of a
$\mathbb{Z}_2$ spin liquid.
Our work suggests a systematic search and classification of classical
spin liquids as a worthwhile endeavour.
\end{abstract}

\maketitle

{\it Fractionalisation} is one of the several
unusual properties generally observed in systems evading low
temperature conventional symmetry breaking ordered states
in favor of unconventional topological orders.
On account of such exotic behavior, much attention has been devoted
to the identification of systems exhibiting such new topological physics. 
Frustrated magnets~\cite{ramirez1994strong,moessner_ramirez,chalker2010geometrically}
have played a prominent role, where several spin liquids (SL)~\cite{anderson1973resonating,fazekas1974on}
starting in the late 90s~\cite{villain1980order} were identified~\cite{moessner1998properties,kitaev2003fault,moessner2001resonating}. 

While by now a multitude of quantum SL have been discovered~\cite{lee2008end} and
classified~\cite{wen2002quantum}, the situation with classical
Heisenberg SL is comparatively much sparser. The first Heisenberg
spin liquid to be identified unambiguously, the antiferromagnet on the
pyrochlore lattice~\cite{moessner1998low,moessner1998properties} is a $U(1)$
spin liquid exhibiting pinch-points in its structure factor indicating
algebraically decaying
correlations~\cite{zinkin1996frustration,moessner1998low,moessner1998properties,isakov2004dipolar,henley2010coulomb},
as well as fractionalisation of its microscopic degrees of freedom:
disorder in the form of dilution creates new, weakly-interacting,
magnetic degrees of freedom which possess a half of the microscopic
magnetic moments of the Heisenberg
model~\cite{henley2001effective,sen2011fractional}.

Such fractionalisation is perhaps the cleanest signature of spin-liquidity
in such a classical setting, as definitions in terms of topological field
theory are frustrated by the bulk gapless excitations due to the continuous
classical nature of the Heisenberg spins.

Given the by now overwhelming variety of known quantum spin
liquids (for an example, see Ref.~\onlinecite{song2015space}), it may
therefore come as a surprise that no corresponding richness appears
to exist for classical Heisenberg magnets: the $U(1)$ case is the only
one studied in detail. It turns up in many settings, such as the
checkerboard and pyrochlore lattices
(for $n\neq2$ component spins)~\cite{moessner1998low,moessner1998properties},
the kagome (for $n>3$ component spins)~\cite{reimers1993order,garanin1999classical},
or the SCGO `pyrochlore slab'~\cite{lee1996isolated}.

\begin{figure}
	\includegraphics[width=\columnwidth]{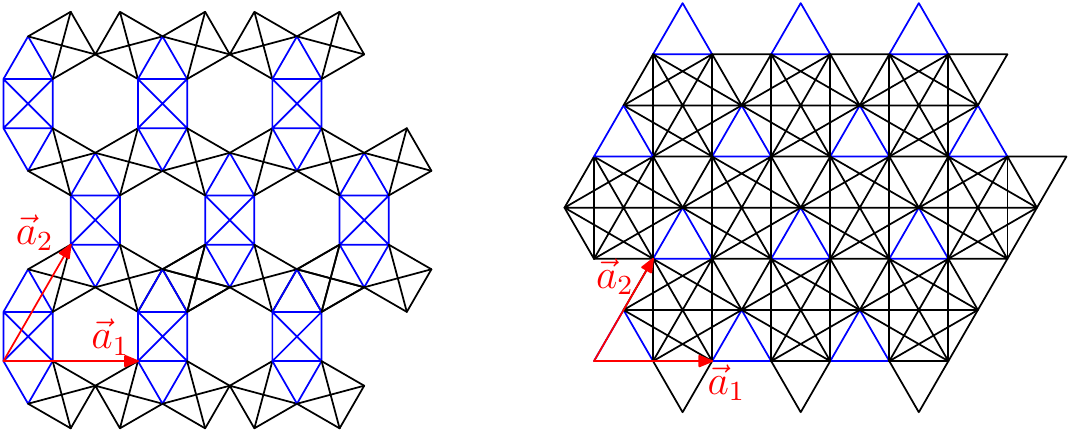}
\caption{Our Heisenberg models are defined on the ruby (left)
  and a kagome (right) lattices. These can be respectively seen
  as fully connected squares forming a kagome, and fully connected
  hexagons forming a triangular, lattice. A choice of basis is
  indicated by the connected sets of blue bonds, and the underlying
  primary vectors of the Bravais lattice in both cases are also
  presented.}
\label{fig:ruby+kag}
\end{figure}

Here we ask the question whether this absence of evidence of other
types of spin liquid is evidence of absence. The answer is that there
is indeed more diversity than has been so far recognised: we identify
a new SL class which does not exhibit algebraic correlations in the
$T\rightarrow0$ limit, and nonetheless displays spin fractionalisation.

In the following, we consider two Heisenberg models, defined on
variants of the $(3,4,6,4)-$Archimedean lattice (known as the ruby
lattice; for a nice introduction to Archimedean lattices see
Ref.~\onlinecite{richter2008quantum}) and of the kagome lattice,
Fig.~\ref{fig:ruby+kag}.
When viewed as corner-sharing networks of clusters, these lattices do not
allow the conventional mapping from spin to flux variables to obtain an
emergent $U(1)$ gauge field~\cite{henley2010coulomb}. We provide the
solution of the corresponding classical $O(n)$ models in the large-$n$
limit and show numerically that this captures well the behaviour of the
Heisenberg $n=3$ model.  These solutions cleanly exhibit the features
mentioned above, including the exponential decay of spin correlations
alongside the appearance of fractional moments in the presence of
dilution with non-magnetic impurities.  We close with a discussion of
the broader picture, in particular exhibiting the connection of this
new SL to a known class of quantum $\mathbb{Z}_2$ SL.

\textit{Model.}---
Practically all Heisenberg models with a SL phase are defined on a
lattice consisting of clusters, such that all pairwise interactions
within a cluster $\alpha$ have equal strength~\cite{henley2010coulomb}.
This implies
\begin{align}
H = J\sum_{\la i,j\ra}\vec{S}_i\cdot\vec{S}_j
  = \frac{J}{2}\sum_{\alpha}(\vec{S}_{\alpha})^2 + \text{const.},
  \label{eq:constr}
\end{align}
with $\vec{S}_{\alpha}=\sum_{i\in\alpha}\vec{S}_i$, the total
spin of a cluster. For an antiferromagnetic
Heisenberg model, any state satisfying the  {\it local constraints} $\vec{S}_{\alpha}=0$ is a ground state.

Lattices of corner-sharing clusters include kagome,
checkerboard, and pyrochlore~\cite{chalker2010geometrically}.
Here,\textit{the clusters themselves occupy a bipartite lattice}.
At least within the limit that spins have an infinite number of
components (the so-called large-$n$ limit~\cite{garanin1999classical};
a finite number of components may, in some cases, lead to an entropic
selection of part of the ground states~\cite{villain1980order,chalker1992hidden,moessner1998low,moessner1998properties},
and to a loss of the low temperature liquid phase),
one finds a classical SL phase with algebraic
correlations at zero temperature and a correlation length that diverges
as $T\rightarrow0$~\cite{henley2010coulomb,isakov2004dipolar}.

Such models have long been studied, 
and the concomitant SL phase has always turned out to host 
an emergent $U(1)$ gauge field
in its Coulomb phase. 
This yields  characteristic {\it pinch-points} in the $T=0$ structure factor~\cite{henley2010coulomb}.

Here, we consider generalizations of such models by identifying cases
where the clusters themselves do {\bf not} occupy a bipartite lattice.
Does this change lead to loss of liquidity fractionalisation and/or pinch points?

Two options in 2d of corner-sharing, \textit{non-bipartite} lattices of
clusters are shown on Fig.~\ref{fig:ruby+kag}.  The left panel illustrates
the ruby lattice, where the clusters (square plaquettes) occupy a
kagome lattice. The right panel in turn illustrates a variant of
the well known kagome lattice, where all spins within a hexagonal plaquette
interact equally with one another forming a corner-sharing network of hexagons.
This we will simply refer to as kagome lattice in the following.

A detailed study of these classical Heisenberg models seems
to be missing in the literature, but the quantum XXZ model on the kagome
lattice considered here has prominently been studied in
Ref.~\onlinecite{balents2002fractionalization}, where the presence of a
$\mathbb{Z}_2$ quantum SL phase was found. Indeed, this quantum SL is
intimately related to the resonating valence bond liquid (RVB) of the 
triangular quantum model,
as it corresponds to a dimer model where each site hosts exactly
three, rather than just a single, dimer~\cite{raman2010quantum}.

\textit{Fractionalisation and liquidity.}---
The most direct way of establishing fractionalisation is to consider the
behaviour of the model under dilution with nonmagnetic ions (vacancies).
Removing all but one spin of a given plaquette, a local paramagnetic moment,
so-called \textit{orphan}~\cite{schiffer1997two}, emerges in the
system, which is robust down to the zero temperature limit. The local moment 
in the models currently
known, is  of size $1/2$~\cite{sen2011fractional}
or $1/3$~\cite{rehn2015classical}.

For the $U(1)$ SLs, an effective theory for such objects yields that,
in the zero temperature limit, they effectively behave as Coulomb
vector charges, since they exhibit a Coulomb interaction with a thermal
screening length, $\xi_{\text{th}}$, that diverges as $1/\sqrt{T}$.~\cite{sen2011fractional,sen2012vacancy}
It is possible to apply the  hybrid field theory developed in these works 
for the orphans surrounded by a bath
of large-$n$ spins to the models considered here.

\begin{figure}[t]
	\includegraphics[width=\columnwidth]{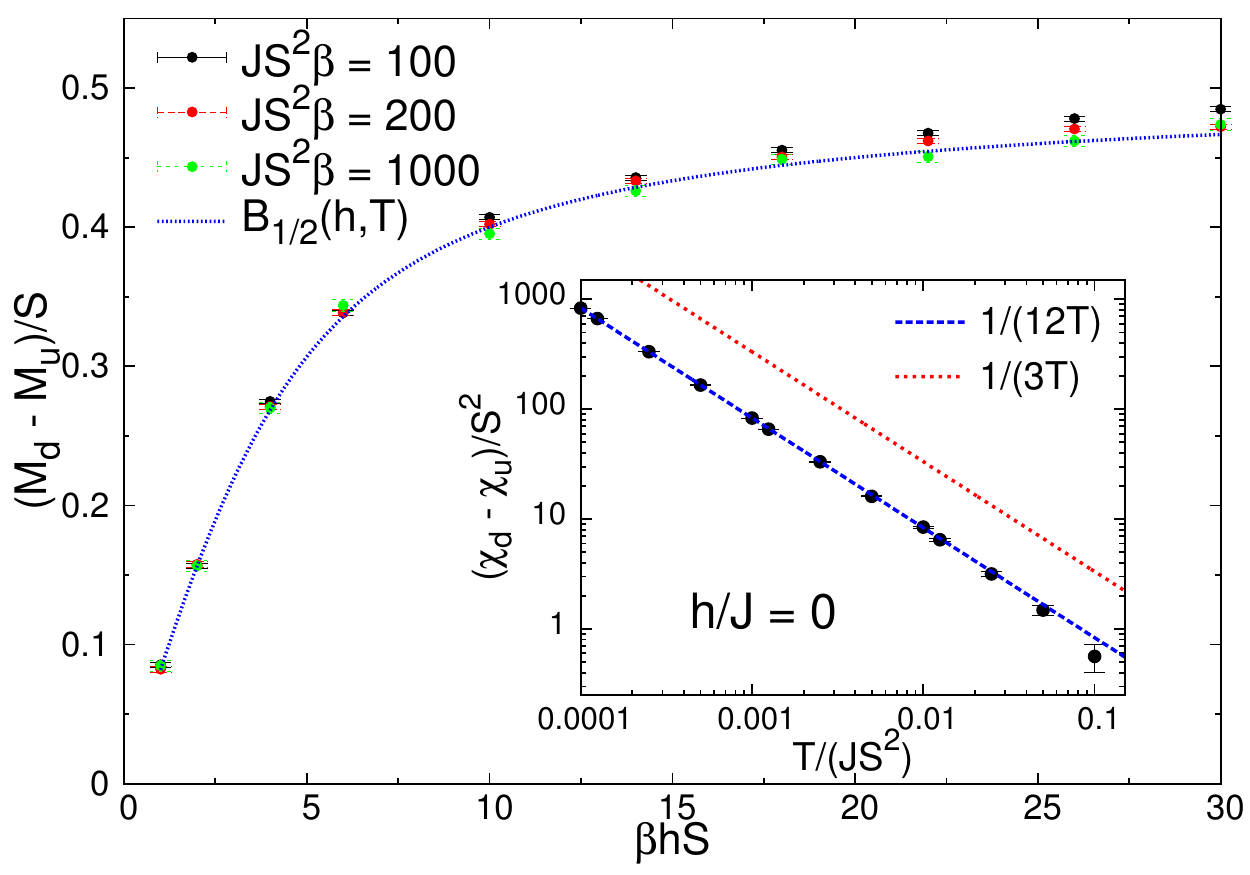}
\caption{The diluted system with an orphan exhibits a fractional moment of
size $S/2$: the total magnetization of a diluted system subtracted from that
of an undiluted one follows the magnetization curve corresponds to a free
fractional moment (main plot), as does the zero-field susceptibility (inset).
Dots denote Monte Carlo simulations of the kagome lattice model;
an analogous result holds for the ruby lattice.}
\label{fig:fract}
\end{figure}

We indeed find that the presence of  fractionalised $1/2$-orphan moments
also occurs in models considered here. Monte Carlo simulations verify
this fact for the Heisenberg case, see Fig.~\ref{fig:fract}.
This is our first central result, as it confirms that low temperature
correlations of these Heisenberg models are not simply those of a
trivial paramagnet, but instead reflect a richer structure, to which
we turn next. 

From the hybrid field theory, one can derive an orphan interaction
in terms of correlators of the pristine system, the so-called
\textit{charge-charge correlations} between the total spin of the
two $\alpha$ clusters located at the respective orphan plaquette
positions,
$\la\vec{S}_\alpha(\vec{r}_1)\cdot\vec{S}_\alpha(\vec{r}_2)\ra$.

\begin{figure}[t]
	\includegraphics[width=\columnwidth]{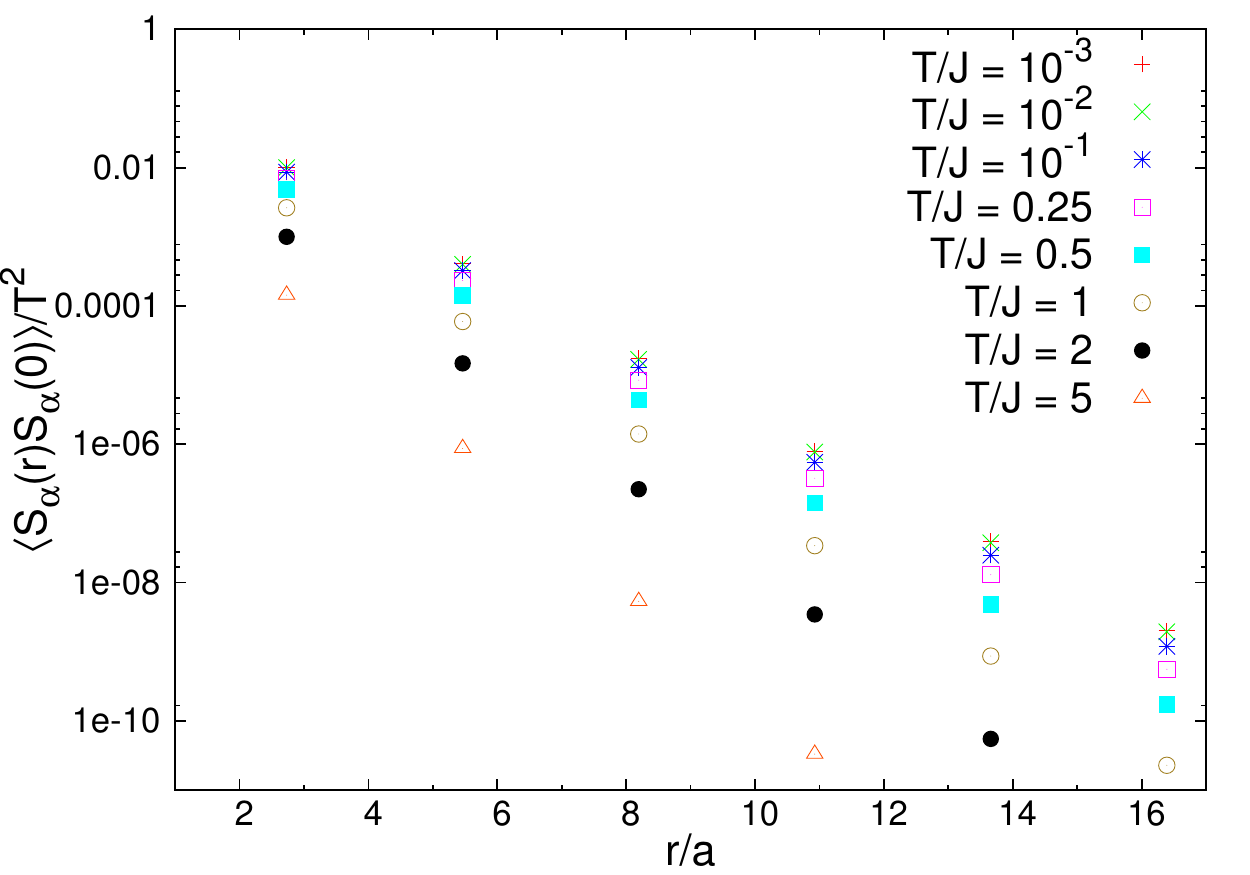}
	\includegraphics[width=\columnwidth]{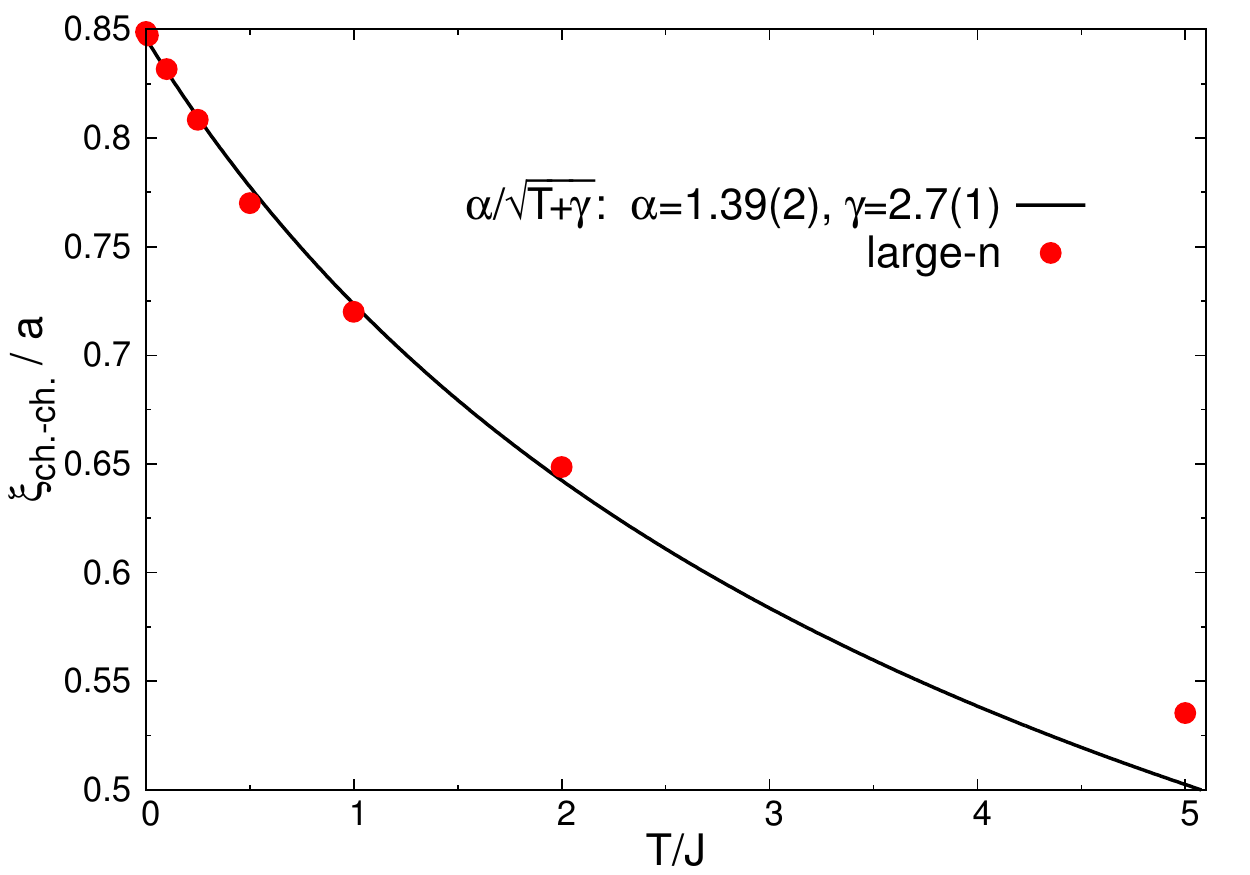}
\caption{The correlation length (bottom) associated to charge-charge
correlations (top) on the ruby lattice, as obtained from a
numerical solution of the large-$n$ equations. The large-$n$ result
for this correlation length (dots, bottom panel) behaves, at low
$T$, as predicted by Eq.~\ref{eq:gapxi} (line).
The length scale $a$ corresponds to the nearest neighbor distance.}
\label{fig:numchchXiVsT}
\end{figure}

We find that these correlators are extremely short ranged,
Fig.~\ref{fig:numchchXiVsT}, with orphan-interactions decaying
exponentially quickly -- rather than algebraically as is the case for
Coulomb orphans. The exponentially decaying orphan-correlations were
also verified directly in Heisenberg Monte Carlo simulations.

This exponential decay results from a feature of the adjacency
matrix spectrum of the lattices studied here, which crucially is
\textit{gapped}. As explained in more details on the Appendix,
this gap replaces the divergent correlation length of the bipartite
U(1) case with
\begin{align}
\xi_{gap}\propto 1/\sqrt{T+\gamma}
\label{eq:gapxi}
\end{align}
with $\gamma>0$. 
We confirm this prediction directly from our numerical solution
of the large-$n$ equations, Fig.~\ref{fig:numchchXiVsT}, which features a
finite correlation length even in the $T\rightarrow0$ limit, smaller
than a nearest neighbor distance, $a$.

\begin{figure*}
\begin{minipage}[t]{.32\textwidth}
  \centering
  \includegraphics[width=1.2\linewidth]{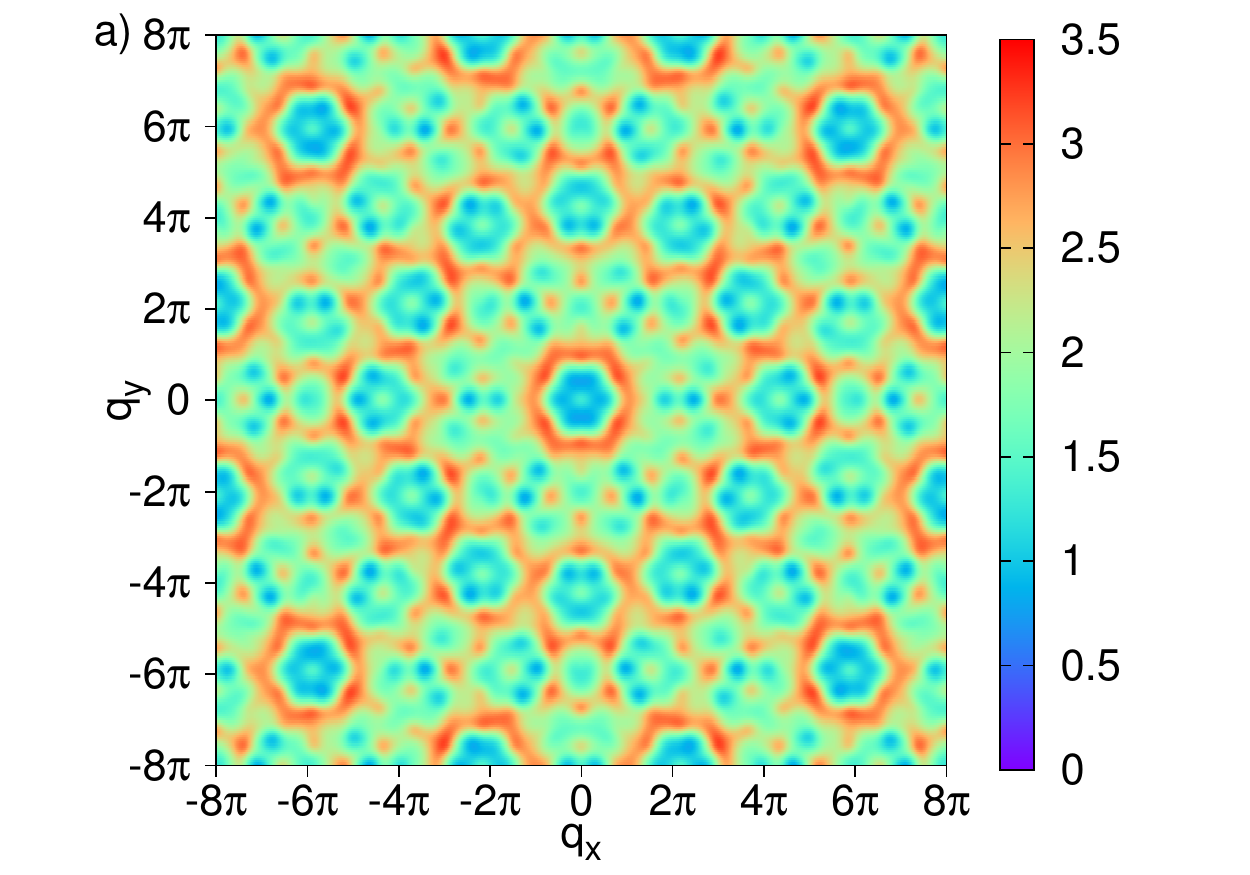}
\end{minipage}\hfill%
\begin{minipage}[t]{.32\textwidth}
  \centering
  \includegraphics[width=1.2\linewidth]{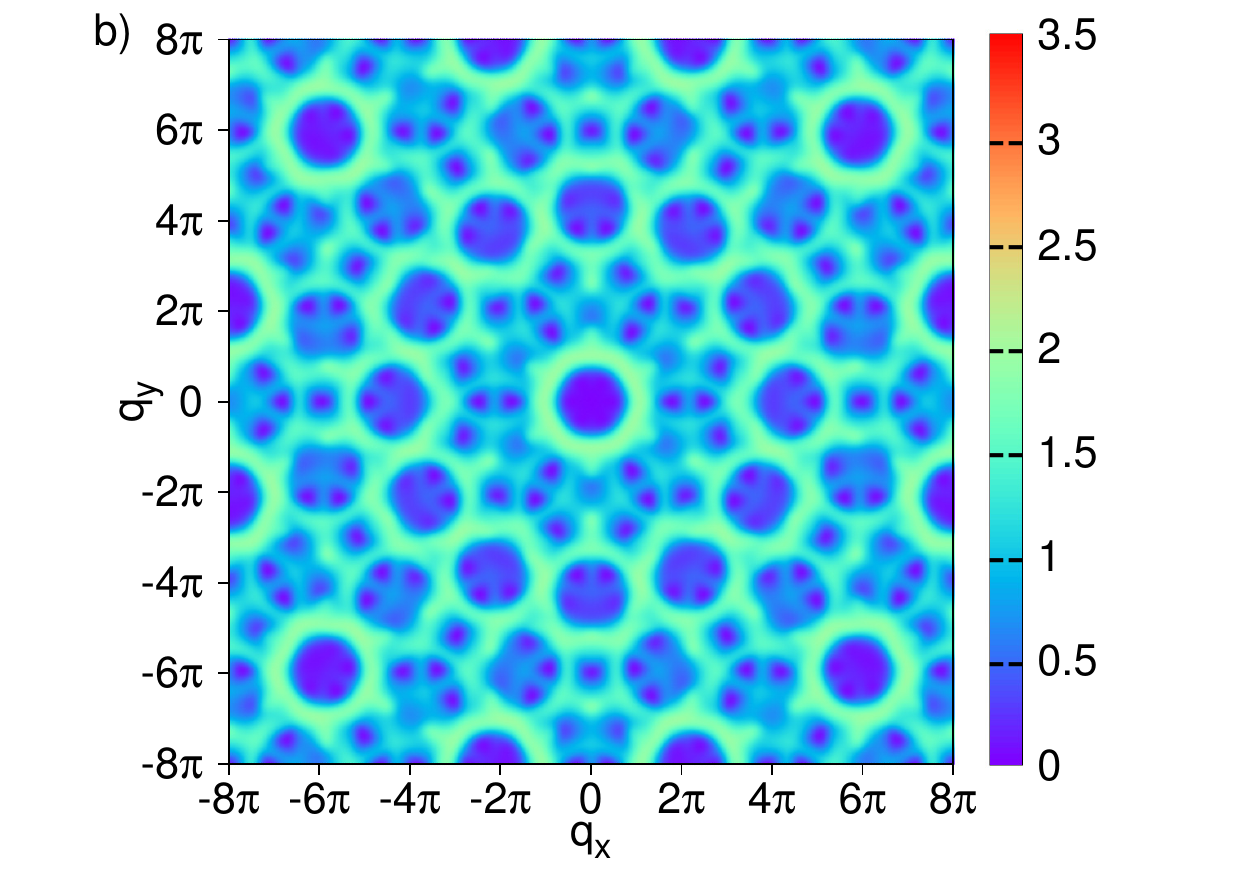}
\end{minipage}\hfill%
\begin{minipage}[t]{.32\textwidth}
  \centering
  \includegraphics[width=1.2\linewidth]{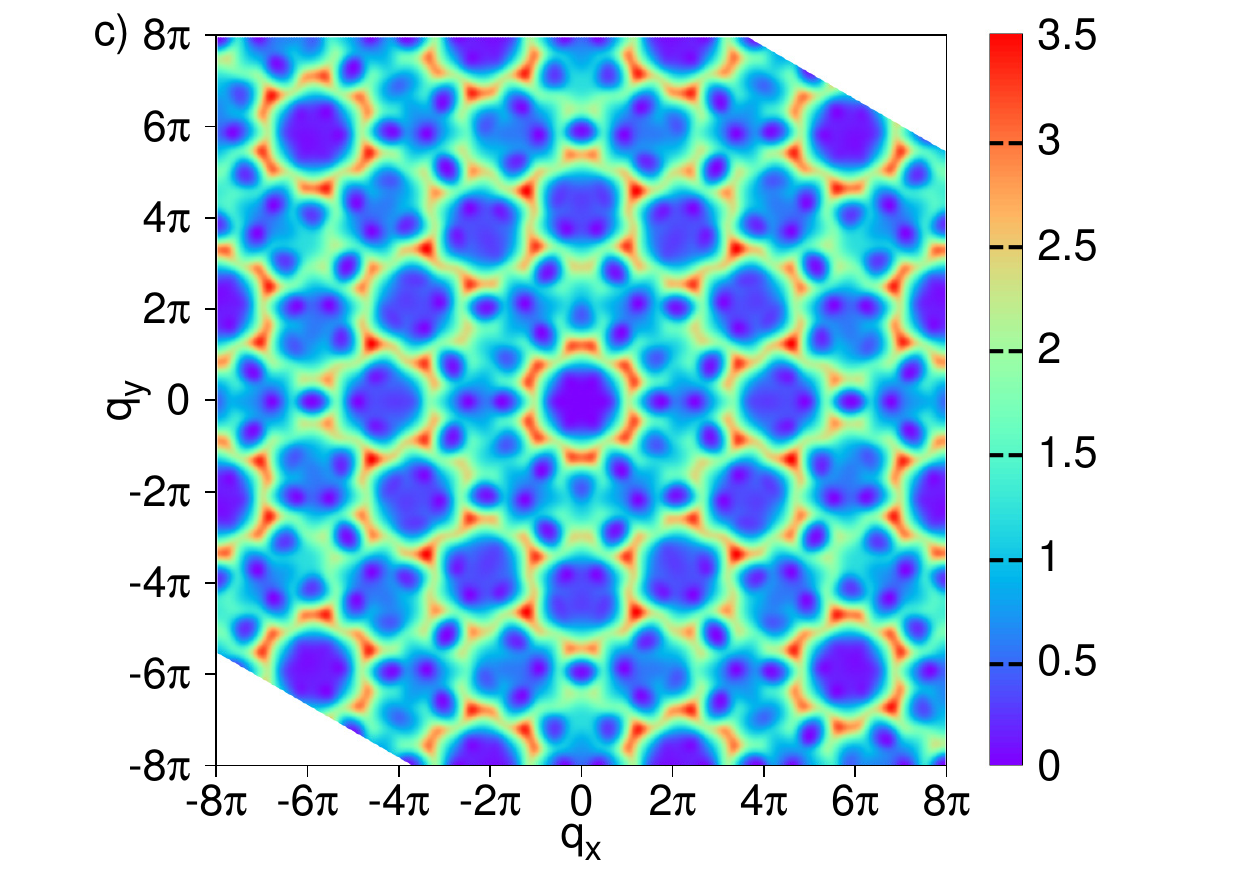}
\end{minipage}
\caption{Spin structure factors on the ruby lattice:
a) analytical result from the large-$n$ approach at zero temperature;
b) Monte Carlo result for a system with Heisenberg spins at inverse
temperature $\beta J = 20$ and $N=7776$ spins ($L=36$); c) result
from a worm Monte Carlo simulation at $T=0$ Ising model, for a
system with $N=10584$ spins ($L=42$).}
\label{fig:sfruby}
\end{figure*}

To further check the absence of any ordering tendency in the Heisenberg
$n=3$ case (due to some order-by-disorder mechanism in the $T\rightarrow 0$ limit), 
we directly study the spin structure factor. The Monte Carlo result
for the ruby lattice Heisenberg model is presented in
Fig.~\ref{fig:sfruby}, obtained from  a combination of parallel
tempering, microcanonical and heat bath moves.
This also displays the  analytical large-$n$ $T=0$ result, as well as
the Ising $n=1$ case at $T=0$ mentioned below  (results for the kagome
case are analogous).
These differ quantitatively, but not qualitatively from each other.

Our simulations reach lattice sizes of $L\times L$ unit cells, with
$L=36$ on the ruby lattice (and $L=24$ on the kagome lattice; not shown),
and the peak heights saturate at large $L$.
This is consistent with quickly decaying correlations
-- the pair spin correlations computed in our simulations are observed
to decay exponentially.

Crucially, and this our next central result, the structure factors do
not present the non-analyticities, such as pinch-points, known to occur
in the $U(1)$ liquids on lattices with corner-sharing structure.

\textit{Discussion.}---
We have presented two Heisenberg models exhibiting
(i) orphan fractionalisation,
(ii) exponentially decaying correlations down to $T\rightarrow0$,
and (iii) absence of order-by-disorder.
Together, these establish the existence of
a new type of classical SL.

In order to embed this in the known lore of spin liquidity, 
let us consider the corresponding
Ising models at zero temperature.
These are related
to dimer coverings on non-bipartite lattices by the following map.

Ising variables sitting on the ruby (kagome) lattice, can also
be seen as variables sitting on the bonds of a kagome (triangular)
lattice. Say that only those bonds with an up spin have a dimer.
Therefore the Ising model ground states are
equivalent to a double dimer covering, or loop model, on the kagome
lattice for the ruby lattice; and to the triple dimer covering of the triangular lattice
mentioned above, for the kagome lattice model.

Such dimer models on non-bipartite lattices often present
only short-range correlations~\cite{raman2010quantum}, which
we have confirmed  by  computing correlations 
 with a worm loop Monte Carlo algorithm. The corresponding
structure factor presented on Fig.~\ref{fig:sfruby} shows qualitative
agreement to the Heisenberg and large-$n$ results. Again, no non-analyticites,
such as pinch-points, are discernible, reflecting the short-range nature of
the spin correlations.

Despite this evidence for absence of ordering, the system is still
not simply in a trivial paramagnetic phase. In fact, an emergent 
$\mathbb{Z}_2$ gauge structure is a well-established possibility for such
dimer models~\cite{raman2010quantum}.
The $\mathbb{Z}_2$ gauge structure arises from the fact that the
set of possible ground states split into winding sectors, such that
local moves within the ground state do not connect configurations with
different winding parities. This is usually seen by considering a non-contractible line
on a torus/cylinder  the system is defined on, and determining the
parity of the number of dimers crossing this line. 
Allowed moves consist of loops visiting
alternately occupied and non occupied
bonds, and exchanging these; such a local rearrangement cannot change the winding parity.

While the $U(1)$ spin liquids on the bipartite lattices of clusters are endowed with a
\textit{winding number} of dimers, the non-bipartite case considered here only allows for a 
 $\mathbb{Z}_2$ winding parity. Thence, by 
analogy to the classical Heisenberg magnet on the
pyrochlore lattice, which retains the $U(1)$ gauge structure of the corresponding 
Ising model (spin ice)~\cite{isakov2004dipolar},
the cases considered here are classical Heisenberg analogies of $\mathbb{Z}_2$ spin liquids.

We believe that the reason their existence has so far been overlooked may have to do with the fact
that non-$U(1)$ frustrated systems in settings considered so far 
have very different Ising and Heisenberg  low-T behaviors, e.g., the Ising triangular antiferromagnet,
 algebraic ground state ensemble, is replaced in the Heisenberg case
by a magnetically ordered ground state.

We have thus shown that classical fractionalisation can occur beyond the $U(1)$ case.
The effect of dilution at low temperatures is therefore to create
very short-ranged interacting spin-texture complexes, which fluctuate as simple
paramagnets with fractional moment of $S/2$.

The necessary conditions for the appearance of such fractional moments are
as yet unknown. More generally, the  abundance of lattice geometries, and the possible influence of further
terms in the Hamiltonian, e.g., anisotropic interactions, remain to be
studied. 
For instance, a  recently
studied SL of Ref.~\onlinecite{benton2010spin}, with anisotropic
interactions on the pyrochlore lattice, exhibits `pinch lines'; while a 
 a new Heisenberg spin liquid on the $J_1-J_2-J_3$ honeycomb
 has been shown to exhibit fractionalised moments of $1/3$, albeit
ultimately also exhibiting pinch-points~\cite{rehn2015classical}.

A general search may thus reveal many further suprises, and a proper classification
of  classical spin-liquid behavior -- known or yet to be discovered --  is a task
that  calls for further research.

\section*{Acknowledgements:}
The work in Dresden was supported by the Deutsche Forschungsgemeinschaft
under grant SFB 1143. The authors thank Kedar Damle for collaboration on
closely related work. The work of AS is partly supported through
the Partner Group program between the Indian Association for the
Cultivation of Science (IACS) and the Max Planck Institute for the
Physics of Complex Systems. JR acknowledges the hospitality of IACS
during the initial stages of this work.

{\it Appendix.}---
Spin correlations at large-$n$  are:
\begin{align}
\la S^{\mu}_{-\vec{q}}S^{\nu}_{\vec{q}}\ra = (\hat{M}^{-1})_{\mu\nu}
\end{align}
where $\mu$, $\nu$ indicate any two atoms in the basis, and the matrix
$\hat{M}$ is related to the interaction matrix, $\hat{V}$, by:
\begin{align}
\hat{M} = \beta J\hat{V} + \lambda \mathbb{1},
\end{align}
with Lagrange multiplier $\lambda$ enforcing  spin
normalization.

Charge-charge correlations,
e.g., in the modified kagome lattice model (where only a single dispersive
band contribute to these correlators), are:
\begin{align}
\la S_{\hexagon}({-\vec{q}})S_{\hexagon}({\vec{q}})\ra
 = \frac{\nu_{kag}T}{\lambda T+\nu_{kag}},
\end{align}
where $\nu_{kag}$ is the eigenvalue of the dispersive band:
\begin{align}
\frac{\nu_{kag}}{J} = 3+(\cos\vec{q}\cdot\vec{a}_1 + \cos\vec{q}\cdot\vec{a}_2 + \cos\vec{q}\cdot\vec{a}_3),
\label{eq:dispeval}
\end{align}
In general, the denominator of the charge-charge correlations
expressions hosts terms of the form $(\lambda'+\nu_i(\vec{q}))$, with
$\lambda' = \lambda / \beta J = \lambda T / J$, and $\nu_i(\vec{q})$
describing the dispersive bands.
This is so, since the matrix inversion involved in computing the
correlations is given by
$G = (\hat{M}^{-1})_{\mu\nu} = \tilde{g}_{\mu\nu}(\vec{q})/\det(\hat{M}(\vec{q}))$,
and in general
\begin{align}
\det\lnb\frac{\hat{M}}{\beta J}\rnb = \lambda'^{n_g}\prod_i(\lambda'+\nu_i(\vec{q})),
\label{eq:detgen}
\end{align}
where $n_g$ denotes the number of ground state flat bands.
The eigenvalues generally depend on a function
$s_{\vec{q}}$ reflecting the symmetries of the lattice at hand.
In the two models considered here, 
\begin{align}
s_{\vec{q}} = \cos{\vec{q}\cdot\vec{a}_1} + \cos{\vec{q}\cdot\vec{a}_2} + \cos{\vec{q}\cdot\vec{a}_3}.
\end{align}
For computing correlations in real space
at large distances, $R\rightarrow\infty$, one can 
expand the symmetric functions $s_{\vec{q}}$ around $\vec{q}=0$;
in our case,
\begin{align}
s_{\vec{q}} \approx -1.5 + ( {q'_x}^2+{q'_y}^2 ),
\end{align}
with $\vec{q}'$ possibly rescaled in relation to $\vec{q}$,
in order to include irrelevant prefactors.
For the case of the kagome charge-charge correlations,
\begin{align}
C = \sum_{BZ}\la S_{\hexagon}({-\vec{q}})S_{\hexagon}({\vec{q}})\ra\exp(i\vec{q}\cdot\hat{r}R) \nn\\
 \sim \iint d^2\vec{q}\exp(i\vec{q}\cdot\hat{r}R) \frac{\nu_{kag}T}{\lambda T+\Delta + ( {q'_x}^2+{q'_y}^2 )}
\end{align}
where $\Delta$ is  related to $\nu_{kag}$, and
hence the band gap. Thus, the correlations
have an asymptotic behavior $\exp(-R/\xi)$, with
\begin{align}
\xi \sim \frac{1}{\sqrt{T + \gamma}} .
\end{align}

\bibliography{ref_Z2}
\end{document}